\documentclass[aps,prl,reprint,letterpaper,twocolumn,10pt,superscriptaddress]{revtex4-1}

\usepackage[utf8]{inputenc}

\usepackage{hyperref} 
\hypersetup{ 
     colorlinks=true, 		
     urlcolor= Blue3,  		
     linkcolor= Blue3, 		
     citecolor=Green4,		
}

\usepackage{amssymb} 		
\usepackage{amsfonts}		
\usepackage{amsmath}        
\usepackage{bm}				
\usepackage{mathrsfs}		
\usepackage{graphicx}		
\usepackage[x11names,svgnames]{xcolor}       
\usepackage{color}      

\begin{document}

\title{Noise feedback in an electronic circuit}
\author{Karl Thibault}
\email{Karl.Thibault@USherbrooke.ca}
\affiliation{Département de physique, Institut quantique, Université de Sherbrooke, Sherbrooke, Québec, J1K 2R1, Canada}
\author{Julien Gabelli}
\email{julien.gabelli@u-psud.fr}
\affiliation{Laboratoire de Physique des Solides, CNRS, Université Paris-Sud, Université Paris-Saclay, 91405 Orsay, France}
\author{Christian Lupien}
\email{Christian.Lupien@USherbrooke.ca}
\affiliation{Département de physique, Institut quantique, Université de Sherbrooke, Sherbrooke, Québec, J1K 2R1, Canada}
\author{Bertrand Reulet}
\email{Bertrand.Reulet@USherbrooke.ca}
\affiliation{Département de physique, Institut quantique, Université de Sherbrooke, Sherbrooke, Québec, J1K 2R1, Canada}

\date{\today}
\begin{abstract}
Electronic circuits are built by combining components with known current/voltage characteristics, which are intrinsic to each component and independent of the rest of the circuit. This approach breaks down for nanostructures placed at ultra-low temperature, a phenomenon referred to as Dynamical Coulomb Blockade, and usually attributed to quantum effects. Here we report similar phenomena on a simple circuit at room temperature, devoid of any quantum features: an avalanche diode in series with a resistor, where the current/voltage characteristics of the diode depends strongly on the value of the resistor. We show that the key ingredient for this is the feedback of the noise of the component on itself through the rest of the circuit. Moreover, we have developed a theory that links transport and noise in the presence of an external electromagnetic environment, which explains very well our experimental results. 
\end{abstract}
\maketitle

\section{I. Introduction}
As the size of a tunnel junction becomes smaller and smaller, its electrostatic capacitance $C$ decreases and the energy $E_C=e^2/C$ to add one single electron becomes more and more relevant. When the energy $k_BT$ associated with thermal fluctuations becomes smaller than $E_C$, the charge in the conductor may freeze leading to the suppression of electronic transport. This phenomenon, called Coulomb Blockade, has been extensively studied in submicron size tunnel junctions placed at very low temperature \cite{fulton_observation_1987}, typically below $\sim$1 K, but is also observed in nanoscale islands at room temperature \cite{beaumont_room_2009}. The total suppression of transport may occur in systems with islands coupled through tunnel junctions where the charge on the island cannot change by less than one electron \cite{korotkov_single-electron_1990,averin_theory_1991,krech_master-equation_1993}, while a partial suppression of transport, usually referred to as Dynamical Coulomb Blockade (DCB), is observed in the absence of a conducting island \cite{delsing_effect_1989}. Since DCB is observed at very low temperature and is associated with the tunneling of single charges, it is usually thought of as a quantum effect. Indeed, DCB is well accounted for by the so-called $P(E)$ theory \cite{devoret_effect_1990,ingold_single_1992} which is based on a quantum treatment of the tunnel junction and the circuit it is connected to, i.e. its electromagnetic environment.

Interestingly, when the size of a conductor decreases until it becomes less than electronic mean free path or the electron-phonon interaction length, it becomes more and more possible for the electrons to cross it without exchanging energy, i.e. transport becomes elastic. This leads to the appearance of shot noise: current fluctuations become voltage dependent \cite{blanter_shot_2000}. Shot noise in mesoscopic conductors is ubiquitous: it comes from the partitioning of the electrons, which are sometimes reflected, sometimes transmitted by the sample, and only channels that are perfectly transparent are devoid of it. Partition noise is described by a binomial process, and thus is not Gaussian, i.e. it has higher order cumulants beyond its variance. The study of such cumulants has revealed that, in a way similar to DCB, the electromagnetic environment of the sample strongly modifies the cumulants of voltage and current fluctuations \cite{reulet_third_2010,kindermann_feedback_2004}.

The link between DCB and environmental effects can be understood in terms of feedback: a current fluctuation $i(t)$ leads to a voltage fluctuation $v(t)$ across the sample because of the finite impedance of its environment. Since noise is voltage dependent, $v(t)$ will modify the probability of subsequent current fluctuations. Links between DCB and shot noise have been carried out in the framework of the $P(E)$ theory \cite{golubev_coulomb_2001,yeyati_direct_2001,kindermann_feedback_2004,safi_one-channel_2004,golubev_electron_2005,parmentier_strong_2011,golubev_coulomb_2012,souquet_dynamical_2013,frey_current_2016}, while another approach based on the noise susceptibility has been proposed \cite{reulet_higher_2005,gabelli_high_2009}. A too na\"ive but elegant and instructive approach consists in describing electron transport in the stationnary state by the probability $P_0(i;V)$ for the instantaneous current to be $i$ when the sample is biased at voltage $V$. In the presence of an external resistor $r$, the voltage across the sample is $v(t)=V-ri(t)$ and the new probability distribution for the current $P(i)$ is given by
\begin{equation}
P(i) \simeq P_0(i;V-ri) \simeq P_0(i;V)-ri \frac{\partial P_0}{\partial V},
\label{eq:P(i)}
\end{equation}
which immediately leads to :
\begin{equation}
I=\langle i\rangle = \langle i\rangle_0-r\frac{\partial \langle i^2 \rangle_0}{\partial V},
\label{eq:I-V_P(i)}
\end{equation}
i.e. a correction to the dc current that involves the derivative of noise with respect to voltage (which should be replaced by the noise susceptibility in a more accurate treatment of the feedback). Furthermore, as sketched in \cite{reulet_higher_2005}, Eq.~(\ref{eq:P(i)}) also leads to the environmental corrections of higher order cumulants of current fluctuations.

From this description of environmental effects it appears that only one ingredient matters for the $I(V)$ characteristic of a component to be affected by what it is connected to: the voltage dependence of the noise it generates. All the electronics industry is based on the idea that the behavior of a component is not affected by the rest of the circuit, so that one can calculate the behavior of the circuit knowing that of the individual components. This is in principle not correct: on the contrary, there is no intrinsic behavior of a component; it always depends on the full circuit. In this article we demonstrate the existence of noise feedback effects analogous to DCB in a circuit at room temperature, an avalanche diode in series with a resistor, and provide the theoretical framework to predict its behavior.

\section{II. Principle of the experiment}
\begin{figure}[b]\centering
\includegraphics[width=\columnwidth]{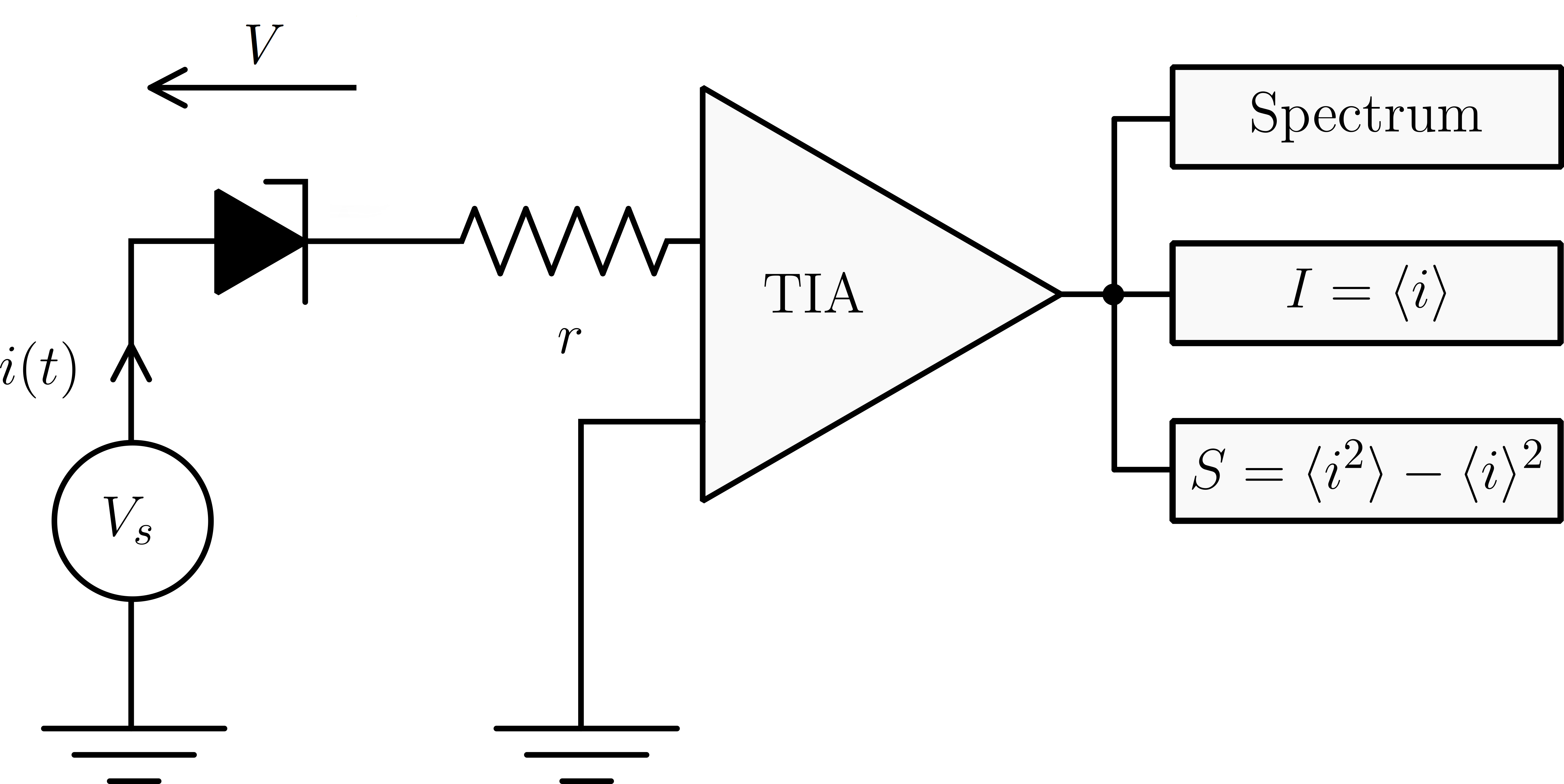}
\caption[]{Experimental setup. TIA represents a transimpedance amplifier.}
\label{fig:setup}
\end{figure}
\begin{figure}[b]\centering
\includegraphics[width=1.05\columnwidth]{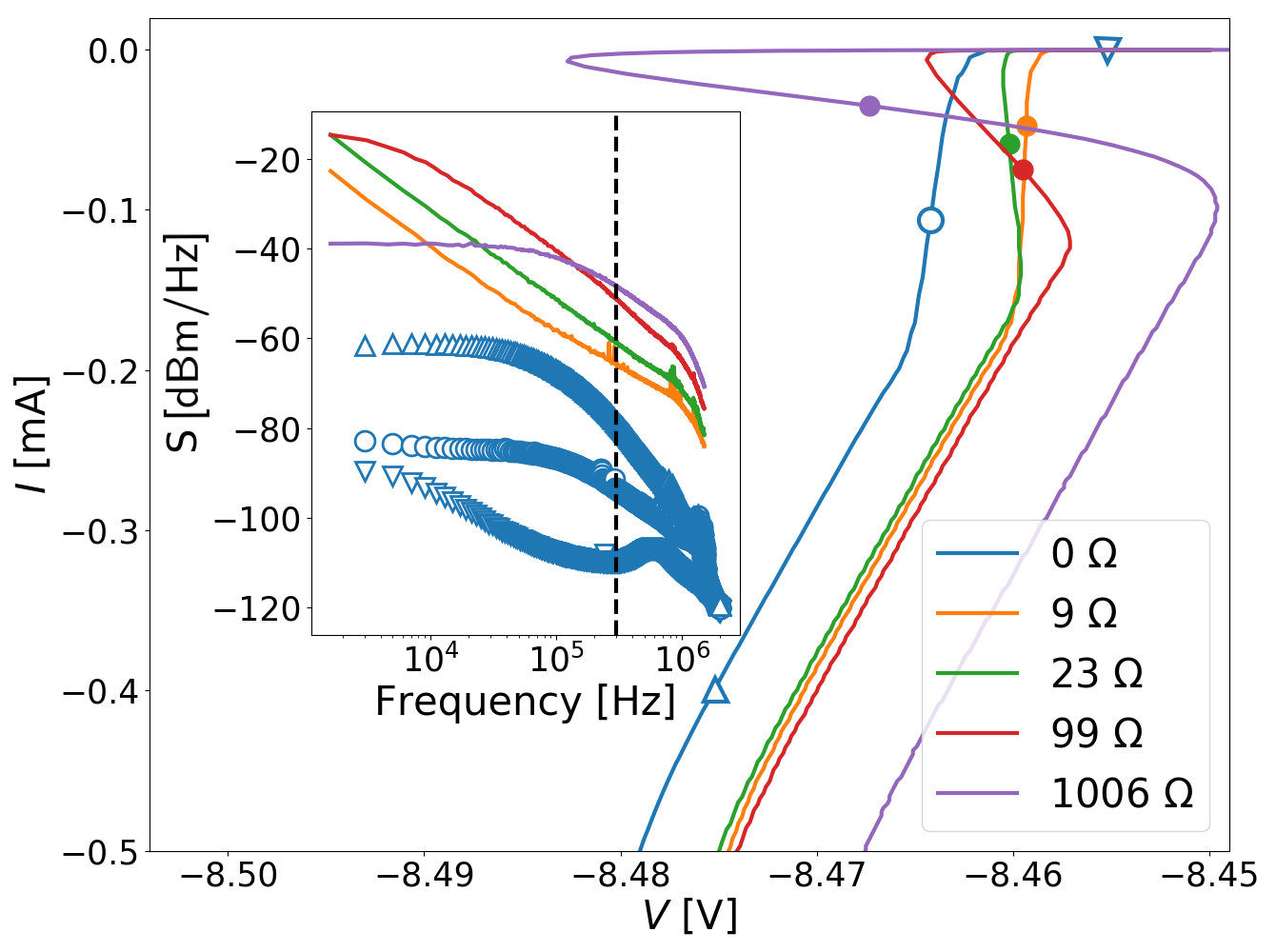}
\caption[]{Current-voltage characteristics $I(V)$ of the avalanche diode for different values of the environmental resistor $r$. \linebreak Inset: Spectra of the current noise of the circuit. Symbols on the $I(V)$ curves represent the voltage point at which the spectra were taken for each resistor. The vertical dashed line at 300~kHz is the upper frequency for the integrated noise measurement $S$.}
\label{fig:I-V_R}
\end{figure}

Our experimental setup is sketched in Fig.~\ref{fig:setup}. We use a commercially available avalanche diode \footnote{\url{http://rf.mrcy.com/RF_Components/Noise_Diodes.html}} usually operated for its noise properties. The avalanche occurs at \mbox{$\approx -8.5~V$} so we will focus our study to voltages nearby. One end of the diode is connected to a voltage source $V_s$, which is shorted by a 10 nF surface-mount ceramic capacitor to ground to ensure a low impedance of the voltage source even at high frequency. The other end of the diode is connected to a thin film surface-mount resistor $r$, which is in turn connected to ground through a transimpedance amplifier. This amplifier plays the role of an ammeter with a bandwidth of 1MHz: it outputs a voltage proportional to the instantaneous current $i(t)$ in the circuit. Then, we use a dc voltmeter to deduce the average current $I=\langle i\rangle$, a spectrum analyzer to measure the noise spectral density of current fluctuations as a function of frequency, and an ac voltmeter to obtain the total variance of current fluctuations $S=\langle i^2(t)\rangle - \langle i\rangle^2$ integrated in the bandwidth 0.1~Hz - 300~kHz. Examples of spectra are given in the inset of Fig.~\ref{fig:I-V_R}. One clearly observes that current fluctuations are well within the bandwidth of the ac voltmeter, so that $S$ can indeed be considered the total noise emitted by the sample. We have also checked that integrating spectra over frequency coincides with the value given by the ac voltmeter.

\begin{figure}[t]\centering
\includegraphics[width=1\columnwidth]{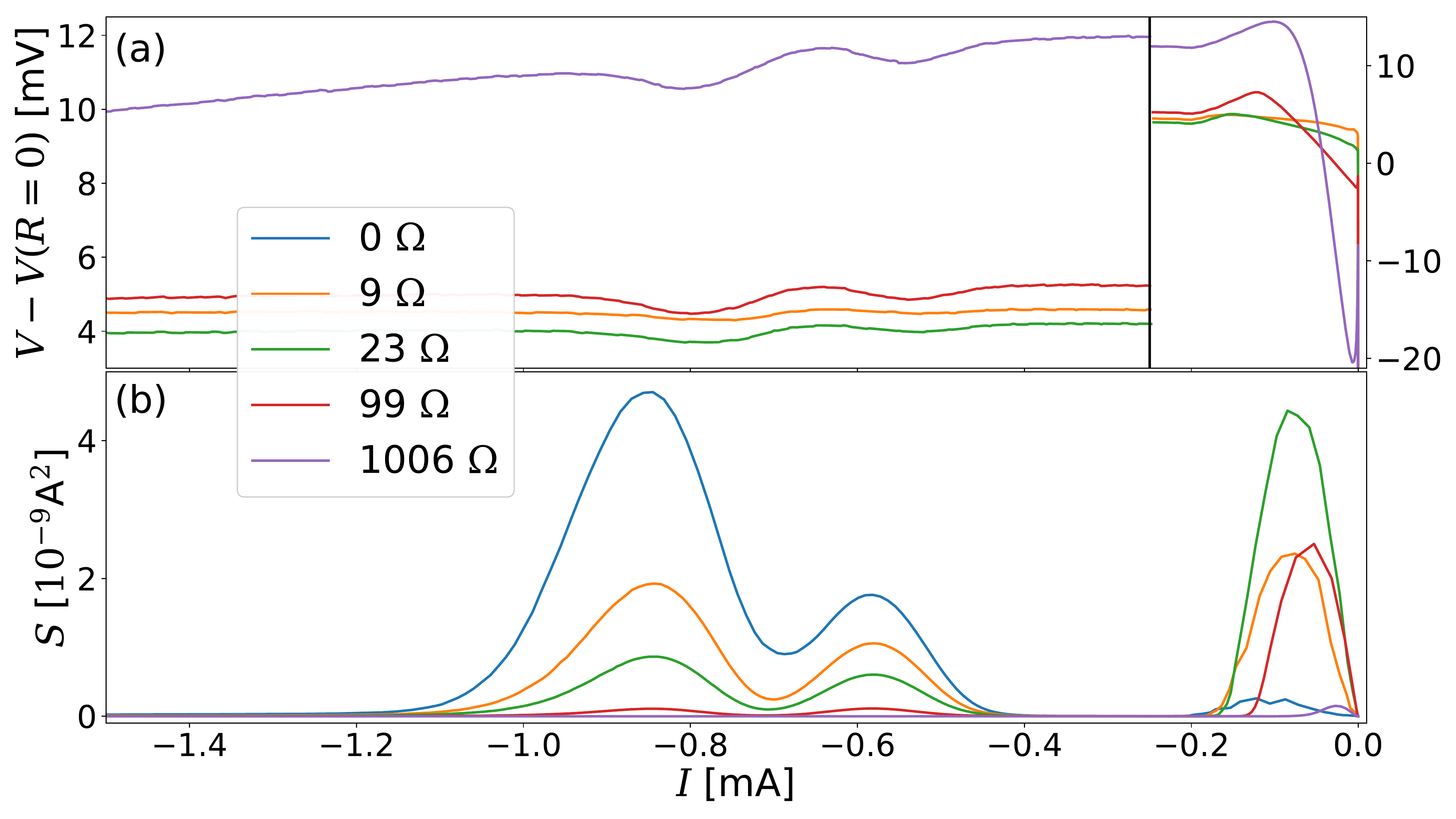}
\caption[]{(a) Voltage $V$ across the diode as a function of the current $I$ for different values of the environmental resistor $r$. Left panel corresponds to voltage between $3\, \mathrm{mV}$ and $13\, \mathrm{mV}$ while right panel corresponds to voltage between $-20\, \mathrm{mV}$ and $15\, \mathrm{mV}$. \linebreak (b) Integrated noise spectral density of the diode as a function of the current $I$ for different values of the environmental resistor $r$.}
\label{fig:V-I_S}
\end{figure}
%

\section{III. Experimental results}
On Fig.~\ref{fig:I-V_R}, we show different current-voltage characteristics $I(V)$ of the avalanche diode for increasing values of the environmental resisto $r$ between $0$ and 1~k$\Omega$ where the voltage is always swept down starting from $-8.45~V$. We note that even at $r=0$, there is a residual impedance seen by the diode of $\approx 5~\Omega$ which consists of the source output and amplifier input impedances. Here $V$ is the voltage across the diode itself, which is calculated using $V=V_s-rI$ ($V_s$ is imposed by the source and $I$ is measured). In the absence of environmental effects, all the curves should be identical, given by the intrinsic characteristic of the diode. We clearly see that it is far from being the case. For $r=0$, the avalanche occurs at $V=-8.462$V and $I$ is a monotonous function of $V$. Starting from $r\gtrapprox 25~\Omega$ the $I(V)$ curve is no longer single-valued at low current, i.e. at the onset of the avalanche. For $r=1$k$\Omega$ the avalanche threshold is pushed down to $-8.487$V, the $I(V)$ characteristic has a large swing and is not single-valued over a broad voltage range. These curves are reminiscent of the $V(I)$ that are predicted for single electron transistor oscillations (SETOs) \cite{ben-jacob_new_1985,averin_coulomb_1986,averin_new_1987,likharev_correlated_1988,negri_charge_2012}. However, we do not observe peaks in the spectra revealing the existence of similar oscillations. Since it is mathematically difficult to deal with non single-valued functions, in the following data analysis we will consider the voltage vs. current $V(I)$ and noise vs. current $S(I)$ characteristics which are single-valued.

Fig.~\ref{fig:V-I_S}(a) shows $V(I)$ up to $|I|=1.5~\mathrm{m}A$ using different scales for high and low currents in order to clearly see the features of both regimes. At high current, all the curves are parallel, but with a clear shift as $r$ is increased. This feature is reminiscent of the shift seen on V-I curves of a tunnel junction with a resistive environment \cite{devoret_effect_1990} known as the Coulomb gap.

To showcase the link between the current dependence of the noise and environmental effects on the transport properties of the avalanche diode, we present $S(I)$ on Fig.~\ref{fig:V-I_S}(b). As $r$ is increased, non-linearities in the $V(I)$ become progressively more apparent around $-0.1~$mA, $-0.6~$mA and $-0.9~$mA, where the $S(I)$ also has its strongest current dependance. However, the value of $r$ does not simply act as a scaling factor. Indeed, the current dependance of $S(I)$ close to the avalanche (${V\sim-0.1}$~mA) is strongest at ${r=23~\Omega}$, even though the non-linearities of $V(I)$ in the avalanche keep increasing with $r$.

In fact, the non-linearity of $S(I)$ around the avalanche seems to be maximal at the exact value of $r$ where $V(I)$ becomes multi-valued. 

Ultimately, even though Eq.~(\ref{eq:P(i)}) and (\ref{eq:I-V_P(i)}) provide a good qualitative picture of the feedback process at cause here, a better model is needed to establish a quantitative link between $S(I)$ and $V(I)$.

\section{IV. Theoretical model}
We have developed a theoretical approach inspired by functional renormalization group theory \cite{gies_introduction_2012,bourbonnais_renormalization_2002,delamotte_hint_2004}. Instead of trying to calculate the characteristics of the device in the presence of an arbitrary external resistor $r$ (current $I(V,r)$ and noise $S(V,r)$), we suppose we know them for a given $r$ and calculate the effect of an additional infinitesimal resistor $\mathrm{d}r$ (see Fig.2 from Supplementary Material at [URL will be inserted by publisher]). We then obtain differential equations with respect to $r$. We model the device by a source of gaussian delta-correlated noise whose variance is voltage-dependent, so that the instantaneous current $i(t)$ in the circuit is given by:
\begin{equation}
i(t)=I(v(t),r)+\xi(t)\sqrt{S(v(t),r)},
\end{equation}
with $\xi(t)$ a gaussian random variable noise of variance 1 and $v(t)$ the instantaneous voltage across the device including the resistance $r$, given by ${v(t) = V_s - i(t)\mathrm{d}r}$. The same $i(t)$ can be seen as that generated by a device that includes a resistance $r+\mathrm{d}r$. Thus it also obeys:
\begin{equation}
i(t)=I(V_s,r+\mathrm{d}r)+\xi(t)\sqrt{S(V_s,r+\mathrm{d}r)}.
\end{equation}
Taking the time average of $i(t)$ and expanding $S(v(t),r)$ to first order in $\mathrm{d}r$ leads to Eq.~(\ref{eq:I-V_P(i)}) with an extra factor $1/2$. This factor was lacking in the same way that a wrong factor appears in the case of a noiseless resistance voltage biased through a resistor as shown in the Supplementary Material at [URL will be inserted by publisher]. Taking the variance of $i(t)$ gives the differential equation on $S$. Higher order cumulants arise from the feedback mechanism induced by the external resistor, but we neglect them and suppose that the device generates Gaussian noise even in the presence of the external resistor. We then can reexpress the equations in terms of current instead of voltage, and introduce the voltage $V$ across the bare device ${V=V_s-rI(V,r)}$. We obtain our main theoretical result:

\begin{align}
\label{eq:dVdiodedR}
\frac{\partial V(I,r)}{\partial r} &= \frac{1}{2}\frac{\partial S(I,r)}{\partial I}\quad \textrm{and}\\
\label{eq:dSdR}
\frac{ \partial S(I,r)}{\partial r} &= \frac{1}{R+r} \left[ -2S(I,r) + \frac{1}{2} \left(\frac{\partial S(I,r)}{\partial I}\right)^2\right],
\end{align}
with ${R(I,r)=\partial V/\partial I}$ the differential resistance of the device. For the full derivation of these equations refer to the Supplementary Material at [URL will be inserted by publisher]. Note that $V$ is the bias on the device (here the avalanche diode) while $S$ is the current noise in the circuit, which is the quantity that we measure. It is related to the intrinsic noise of the device $S_{int}$, which cannot be measured in the presence of the external resistor, by ${S = S_{int} \left(\frac{R}{R+r}\right)^{2}}$. 

\begin{figure*}[th]\centering
\includegraphics[width=2\columnwidth]{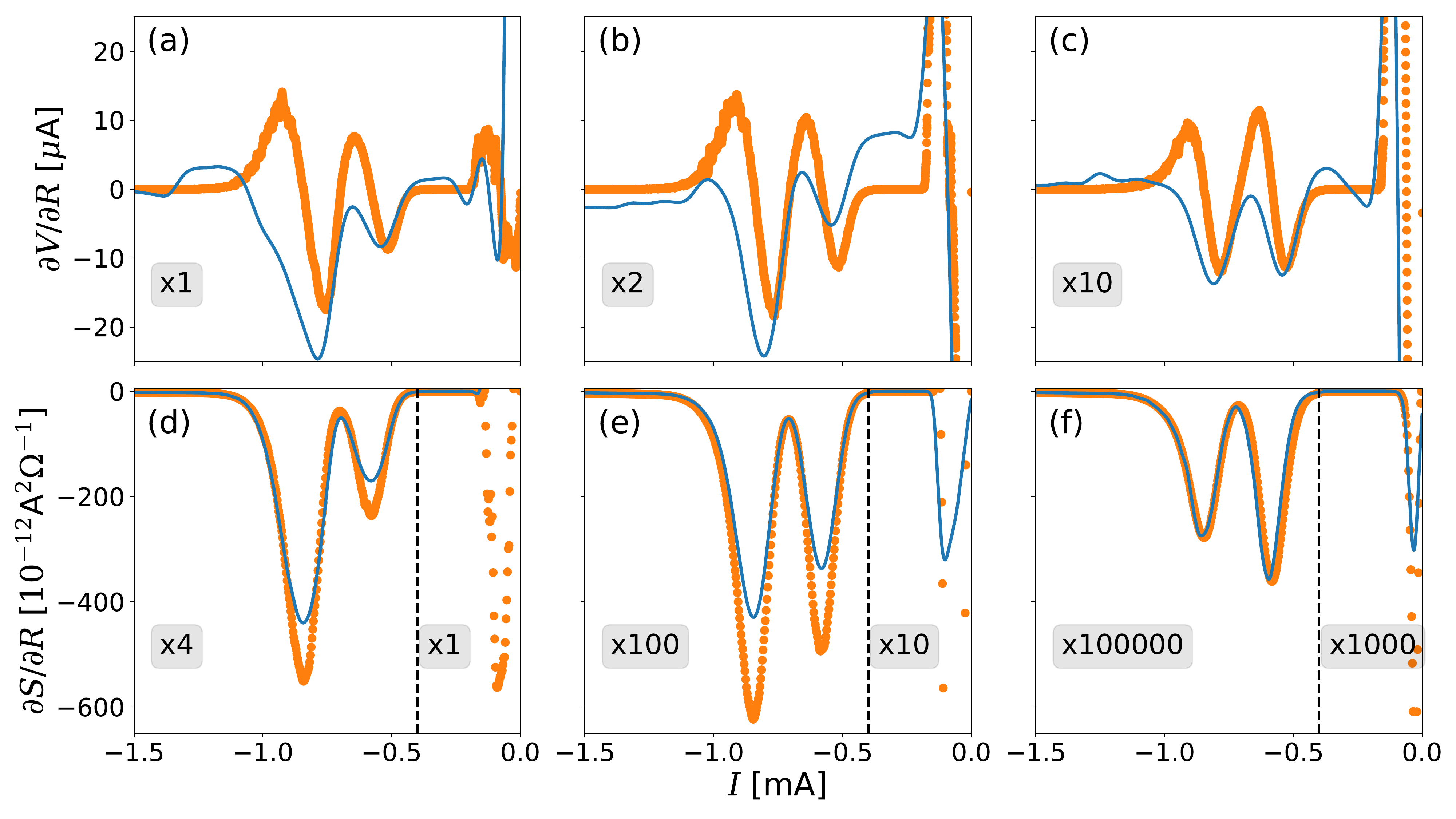}
\caption{Experimental validation of Eq.~(\ref{eq:dVdiodedR}) (top row) and Eq.~(\ref{eq:dSdR}) (bottom row) for different values of resistances. Blue lines and orange dots respectively represent the lhs and rhs of both equations. The values of $(r_1,r_2)$ are as follows: \textbf{(a)} $(2~\Omega, 5~\Omega)$, \textbf{(b)} $(10~\Omega, 15~\Omega)$, \textbf{(c)} $(51~\Omega, 99~\Omega)$, \textbf{(d)} $(10~\Omega, 15~\Omega)$, \textbf{(e)} $(51~\Omega, 99~\Omega)$ and \textbf{(f)} $(1006~\Omega, 1020~\Omega)$. The experimental data are multiplied by a mutiplicative factor in order to keep the y-axis constant along each row.}
\label{fig:dVdR_dSdR}
\end{figure*}

To clarify the meaning of Eqs.~(\ref{eq:dVdiodedR},\ref{eq:dSdR}), it is helpful to look at a simple examples. First, Eq.~(\ref{eq:dVdiodedR}) implies that the $V(I)$ characteristics of the device is not affected by its environment if and only if its noise is current-independent. Let us consider a resistor $R$ in parallel with a capacitor $C$ (in order to keep the bandwidth finite), which generates thermal noise ${S_0=4k_BT/(R^2C)}$. Integrating Eq.~(\ref{eq:dSdR}) with the condition ${S(I,r=0)=S_0}$ leads to ${S=S_0\left(\frac{R}{R+r}\right)^{2}}$, which simply means that the noise generated by $R$ is split between $r$ and $R$. We now turn to the case of a tunnel junction at zero temperature, which obeys: ${V(I,r=0)=RI}$, ${S(I,r=0)=2eB|I|}$ with ${B=1/(RC)}$ the bandwidth of the shot noise emitted by the junction of geometric capacitance $C$. By integrating Eqs.~(\ref{eq:dVdiodedR}) and (\ref{eq:dSdR}) we find ${V(I,r)=RI+\Delta(r) sgn(I)}$ with ${sgn(I)=1}$ for ${I>0}$ and ${sgn(I)=-1}$ for ${I<0}$. A discontinuity appears at ${I=0}$ in the $V(I)$ characteristics: there is no current at low voltage ${|V|<\Delta}$. This corresponds to the existence of a Coulomb gap ${\Delta=(e/C) r/(r+R)}$ which tends to the usual result $e/C$ when ${r\rightarrow\infty}$ \cite{devoret_effect_1990}. For the noise, by integrating Eq.~(\ref{eq:dVdiodedR}) and using the continuity of $S(I,r)$ for $I=0$, we find the intrinsic noise of the device $ S_{int}=S \left(\frac{R+r}{R}\right)^{2}=2eB|I|$. We recover the fluctuation dissipation relations at zero frequency of a tunnel junction characterized by a Fano factor equal to unity in agreement with quantum theories taken at zero frequency \cite{parlavecchio_fluctuation-dissipation_2015,roussel_perturbative_2016}.

Despite its simplicity, our theoretical approach clearly captures the physics of the usual dynamical Coulomb blockade observed in mesoscopic conductors at low temperature. It may give insights into regimes beyond the usual $P(E)$ theory and can be applied to any system regardless of its microscopic description. It has, however, obvious limitations. In particular, the frequency dependence of the noise, impedance and noise susceptibility \cite{gabelli_noise_2008} are all disregarded. Taking into account such a frequency dependence should be possible but is beyond the scope of the present work.

\section{V. Validation}
To validate our theory, we have measured $V(I,r)$ and $S(I,r)$ as described above and checked that Eqs.~(\ref{eq:dVdiodedR},\ref{eq:dSdR}) reproduce the experimental data. Derivatives with respect to current appearing in the equations can be accurately evaluated thanks to the many values of voltage/current measured. Changing the resistances is much more time consuming, so the derivatives with respect to $r$ are replaced experimentally by finite differences. For data measured at two resistances $r_1$ and $r_2$, Eq.~(\ref{eq:dVdiodedR}) is replaced by 
\begin{equation}
\frac{V(I,r_1)-V(I,r_2)}{r_2-r_1}=\frac{1}{2}\frac{\partial \left[\frac{S(I,r_1)+S(I,r_2)}{2}\right]}{\partial I}. \nonumber
\end{equation}
The top row of Fig.~\ref{fig:dVdR_dSdR} shows the experimental verification of Eq.~(\ref{eq:dVdiodedR}), i.e. the effect of the external resistor on the $V(I)$ characteristics, for ${(r_1,r_2)=(2~\Omega,5~\Omega)}$, $(10~\Omega,15~\Omega)$ and $(51~\Omega,99~\Omega)$. The bottom row of Fig.~\ref{fig:dVdR_dSdR} shows the experimental verification of Eq.~(\ref{eq:dSdR}), i.e. the effect of the external resistor on the noise, for ${(r_1,r_2)=(10~\Omega,15~\Omega)}$, $(51~\Omega,99~\Omega)$ as well as $(1006~\Omega,1020~\Omega)$. Clearly, our theoretical predictions of Eqs.~(\ref{eq:dVdiodedR}) and (\ref{eq:dSdR}) are in good agreement with our experimental data over a large range of resistances (from a few Ohms to a kilo-Ohm) and for signal amplitudes spanning more than 5 orders of magnitude. All the oscillations present in the current dependence of the noise lead to oscillations in the voltage exactly at the position we predict, and the overall amplitude of the data is well accounted for. Precise prediction of the amplitude of the peaks is lacking, but we think this is due to the difficulty of measuring derivatives with respect to the environmental resistance $r$. Indeed, the agreement between our theory and experimental data is virtually perfect for ${(r_1,r_2)=(1006~\Omega,1020~\Omega)}$ on Fig.~\ref{fig:dVdR_dSdR} (f), where the finite difference correctly approximates a derivative: ${(r_2-r_1)\ll r_1,r_2}$.

\section{VI. Conclusion}
Electronic circuits are built by assembling components assuming these have a well defined, intrinsic I(V) characteristics. We have shown experimentally that this is incorrect as soon as components exhibit a voltage-dependent noise. We have derived two equations which dictate how a device's intrinsic $V(I)$ and $S(I)$ characteristics change when it is inserted in a resistive electromagnetic environment. The only necessary ingredient is for the device to have current- or voltage-dependent noise. The origin of this dependence can be quantum, as in the case of shot noise and DCB, or not, as shown here in the avalanche diode. We have demonstrated experimentally the relevance of our theory by showing how the intrinsic characteristics of an avalanche diode dramatically depends on its environmental impedance. They show that the idea that components in a circuit behave according to their individual intrinsic $I(V)$ characteristics is erroneous. On the contrary, electronic components interact through their noise, which may lead to strong deviations of their characteristics, as we have observed.
Our results are very general and do apply to a large class of devices. Since almost all devices exhibit voltage-dependent noise, the effects we have described should be ubiquitous in electronic circuits. While we have, for the sake of simplicity, diregarded any frequency dependence (of the impedances, the noise and noise susceptibility), this could be restored with some effort.

\section{Acknowledgments}
We acknowledge fruitful discussions with E. Pinsolle and F. Pistolesi. This work was supported by ANR-11-JS04-006-01, the Canada Excellence Research Chairs program, the NSERC, the MEI, the FRQNT via the INTRIQ, the Université de Sherbrooke via the EPIQ, the Canada First Research Excellence Fund and the Canada Foundation for Innovation.

\bibliographystyle{apsrev4-1}
\bibliography{biblio_vF}

\end{document}


\title{Supplementary material: Noise feedback in an electronic circuit}
\author{Karl Thibault}
\email{Karl.Thibault@USherbrooke.ca}
\affiliation{Département de physique, Institut quantique, Université de Sherbrooke, Sherbrooke, Québec, J1K 2R1, Canada}
\author{Julien Gabelli}
\email{julien.gabelli@u-psud.fr}
\affiliation{Laboratoire de Physique des Solides, CNRS, Université Paris-Sud, Université Paris-Saclay, 91405 Orsay, France}
\author{Christian Lupien}
\email{Christian.Lupien@USherbrooke.ca}
\affiliation{Département de physique, Institut quantique, Université de Sherbrooke, Sherbrooke, Québec, J1K 2R1, Canada}
\author{Bertrand Reulet}
\email{Bertrand.Reulet@USherbrooke.ca}
\affiliation{Département de physique, Institut quantique, Université de Sherbrooke, Sherbrooke, Québec, J1K 2R1, Canada}

\date{\today}

\maketitle

\section{Noise Feedback Theoretical Models}

\subsection{Erroneousness of the first order corrections in toy models}
\begin{figure}[h!]
\includegraphics[height=7.0\baselineskip]{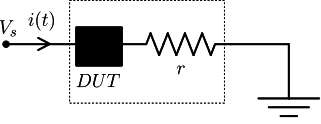}
\caption{Model of a device under test (DUT) without fluctuations.}
\label{fig:prob_montage}
\end{figure}

For the circuit shown on Fig.\ref{fig:prob_montage}, it is tempting to express the noise feedback effects using the probability distribution of current $P(i;V_s)$ through
\begin{equation}
    P(i;V_s) = P_0(i;V_s-ri).
\end{equation}
Let us consider the case of a resistor $R$ with no fluctuations as DUT; we obtain
\begin{equation}
    P_0(i;V_s) = \delta (i-V_s/R).
\end{equation}
The current in the presence of $r$ is then given by
\begin{align}
    \langle i \rangle &= \int i P(i;V_s) \mathrm{d}i \nonumber \\
    &= \int P_0(i;V_s-ri) \mathrm{d}i \nonumber \\
    &= \int i \delta \left(i-\frac{1}{R}(V_s-ri)\right) \mathrm{d}i \nonumber \\
    &= \int \frac{x}{1+r/R} \frac{\mathrm{d}x}{1+r/R} \delta\left(x-V_s/R\right), \quad \mathrm{where} \: x = i(1+r/R) \nonumber \\
    &= \frac{V_s/R}{(1+r/R)^2} \nonumber \\
    &\simeq \frac{V_s}{R}(1-2r/R).
\end{align}
This result is wrong by a factor of $1/2$ compared to the correct result, $\frac{V_s}{R}(1-r/R)$, obtained using basic circuit theory. This offset is the same as the one obtained in Eq.(2) in the main article. This is due to this method's rough use of i as no longer strictly speaking the stochastic variable defined in $P_0(i;V_s)$ (which depends on the constant parameter $V_s$), but as a variable in the usual sense, which is not mathematically correct. However, this faux-pas enables rapid understanding of the underlying physics of the phenomena at play.

\subsection{Renormalization method}
The goal of the model developed herein is to calculate the effect of an infinitesimal change of the environmental resistor ($\mathrm{d}r$) on the characteristics of the device. This technique leads to the equations (5) and (6) in the main article.

This supplementary material expands the derivation of those equations and clarifies the limits of the calculation. See \cite{thibault_effets_2019} for the full derivations.

\subsubsection{Toy-model : no fluctuations}

\begin{figure}[h!]
\includegraphics[height=7.0\baselineskip]{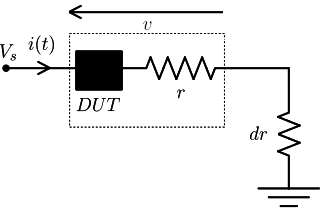}
\caption{Model of a device under test (DUT) without fluctuations considering a change $\mathrm{d}r$ in it's environmental impedance.}
\label{fig:renorm_montage_ss_dI}
\end{figure}

As a first step we suppose that neither the external resistor nor the device we consider exhibit noise, so current and voltage are totally time-independent. We note $I_0(V_s)$ when it is completely isolated, i.e. it's environmental impedance is effectively zero, and polarized by a perfect voltage source $V_s$. When this device is in series with a resistor, it is intead caracterized by $I(V_s,r)$, with $I(V_s,0) = I_0(V_s)$. In the experiments shown in the main paper, this device is an avalanche diode, but it could be anything.

We now define the voltage $v$ as the voltage accross the device and the resistor (see Fig.\ref{fig:renorm_montage_ss_dI}) and look at the effects of an increase in the value of the resistor by ($\mathrm{d} r$).

We have 
\begin{align}
    v &= V_s - I(v,r)\mathrm{d} r \quad \mathrm{and} \\
    I(v,r) &= I(V_s,r + \mathrm{d} r)
\end{align}
so that a Taylor series expansion leads to
\begin{equation}
    I(v,r) = I(V_s,r) - \mathrm{d} r I(v,r) \left(\frac{\partial V_s}{\partial v}\frac{\partial I(v,r)}{\partial V_s}\right)\bigg|_{v = V_s} + O(\mathrm{d} r^2).
\end{equation}
with $\frac{\partial V_s}{\partial v}=1+\frac{\partial I(v,r)}{\partial v}\mathrm{d} r$. Then, taking the limit $\mathrm{d} r \rightarrow 0$ with the fact that $I(v,r) = I(V_s,r + \mathrm{d} r)$ leads to
\begin{align}
    \label{eq:dI/dR_ss_fluct}
    \lim_{\mathrm{d} r \rightarrow 0} \frac{I(V_s,r+\mathrm{d} r) - I(V_s,r)}{\mathrm{d} r} &= \lim_{\mathrm{d} r \rightarrow 0} -I(v,r)\frac{\partial I(v,r)}{\partial V_s}\bigg|_{v = V_s} \nonumber\\
    \Rightarrow \frac{\partial I(V_s,r)}{\partial r} &= -I(V_s,r) \frac{\partial I(V_s,r)}{\partial V_s},
\end{align}

This equation is indeed valid for a simple resistor $I_0=V_s / R_0$, where both sides are equal to $-V_s/R^2$. This result can also be rewritten as $\frac{\partial I}{\partial r} = - \frac{1}{2}\frac{\partial I^2}{\partial V_s}$, which, as mentioned before, is a factor $1/2$ off from toy-model predictions in Eq.(2) from the main article.

\subsubsection{Taking into account voltage fluctuations}
In a real device, voltage fluctuations play a role in the transport mecanisms. Let us now consider the effect of such fluctuations on our calculation.
\begin{figure}[h!]
\includegraphics[height=7.0\baselineskip]{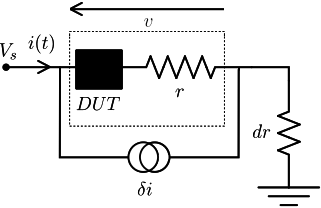}
\caption{Model of a device under test (DUT) with fluctuations.}
\label{fig:montage_complet}
\end{figure}

Suppose that the device exhibits current noise $\langle \delta I_0(V_s)^2 \rangle = S_0(V_s)$, while neglecting fluctuations generated by the external resistor. The instantaneous current in the circuit now becomes time-dependent:
\begin{equation}
    i(t,V_s,r) = I(V_s,r) + f(V_s,r) \epsilon(t),
\end{equation}
with $\langle i(t,V_s,r) \rangle = I(V_s,r)$ and $\delta i(t,V_s,r) = f(V_s,r)\epsilon(t)$, $\epsilon(t)$ being a random variable. These are linked to the noise of the circuit by $S(V_s,r) =  f(V_s,r)^2 \langle \epsilon(t)^2 \rangle$.
The same calculations as in the noiseless case lead to
\begin{equation}
    \frac{\partial i(t,V_s,r)}{\partial r} = -i(t,V_s,r) \frac{\partial i(t,V_s,r)}{\partial V_s},
\end{equation}
which is the same equation for instaneous current as in the noiseless case (Eq. (\ref{eq:dI/dR_ss_fluct})).

However, the mean current is given by
\begin{align}
    \label{eq:renorm_dI/dR}
    \langle i(t,V_s,r+\mathrm{d} r) \rangle &= I(V_s,r+\mathrm{d} r) \nonumber\\
    &= I(V_s,r) - I(V_s,r) \frac{\partial I(V_s,r)}{\partial V_s}\mathrm{d} r \nonumber\\
    \quad &-\langle f(V_s,r) \frac{\partial f(V_s,r)}{\partial V_s}\mathrm{d} r \epsilon(t)^2 \rangle + O(\mathrm{d}r^2) \nonumber\\
    \Rightarrow \frac{\partial I(V_s,r)}{\partial r} &= -I(V_s,r) \frac{\partial I(V_s,r)}{\partial V_s} - \frac{1}{2}\frac{\partial S(V_s,r)}{\partial V_s},
\end{align}
where an addtional term due to fluctuations is now apparent.

It is also possible to calculate an equivalent equation for the noise
\begin{align}
    \delta i(t,V_s,r+\mathrm{d} r) &= i(t,V_s,r+\mathrm{d} r) - \langle i(t,V_s,r+\mathrm{d} r) \rangle \nonumber\\ 
    &= f(V_s,r)\epsilon(t) \nonumber\\
    &\quad - \left[ \frac{\partial f(V_s,r)}{\partial V_s} I(V_s,r)\epsilon(t) +\frac{\partial I(V_s,r)}{\partial V_s}f(V_s,r) \epsilon(t) + f(V_s,r)\frac{\partial f(V_s,r)}{\partial V_s} \left( \epsilon^2(t)-1 \right) \right] \mathrm{d} r.
\end{align}
Thus,
\begin{align}
    S(V_s,r+\mathrm{d} r) &= \langle \delta i(t,V_s,r+\mathrm{d} r)^2 \rangle  \nonumber\\
    &= \langle f(V_s,r)^2\epsilon(t)^2\rangle \nonumber\\
    &\quad - \langle \left[2  f(V_s,r)\frac{\partial f(V_s,r)}{\partial V_s}I(V_s,r) \epsilon(t)^2 R + 2 f(V_s,r)^2\frac{\partial I(V_s,r)}{\partial V_s}\epsilon(t)^2 + f(V_s,r)^2 \frac{\partial f(V_s,r)}{\partial V_s} \left( \epsilon^3(t)-\epsilon(t) \right) \right]\rangle \mathrm{d} r \nonumber\\
    &\quad + O(\mathrm{d} r^2)\nonumber\\
    &=  S(V_s,r) - I(V_s,r) \frac{\partial S(V_s,r)}{\partial V_s}  \mathrm{d} r - 2 S(V_s,r) \frac{\partial I(V_s,r)}{\partial V_s}  \mathrm{d} r + O(\mathrm{d} r^2) \nonumber\\
    \Rightarrow \frac{\partial S(V_s,r)}{\partial r} &= -2 S(V_s,r) \frac{\partial I(V_s,r)}{\partial V_s} - I \frac{\partial S(V_s,r)}{\partial V_s}.
    \label{eq:renorm_dS/dR}
\end{align}
It is noteworthy to mention that the random variable $\epsilon(t)$ has to be gaussian for this result to be valid, otherwise the term $\propto \epsilon(t)^3$ would need to be taken into account, although the magnitudes of corrections from higher cumulants are usually negligible.

At this point, all characteristics are still dependent on the voltage of the source $V_s$. We now define voltage $V$ as the voltage accross the device itself, and recalculate the equations above in terms of this quantity. We then obtain
\begin{equation}
    \frac{\partial I(V,r)}{\partial r}\Bigr|_{V} = -\frac{1}{2} \frac{\partial S(V,r)}{\partial V}\Bigr|_r,
    \label{eq:renorm_dI/dR_Vint_complet}
\end{equation}
\begin{equation}
    \frac{\partial S(V,r)}{\partial r}\Bigr|_{V}= -2 S \frac{G}{1+rG} - \frac{1}{2} \frac{r}{1+rG} \left( \frac{\partial S(V,r)}{\partial V}\Bigr|_r \right)^2,
    \label{eq:renorm_dS/dR_Vint_complet}
\end{equation}
with $G =\frac{\partial I}{\partial V}$ the conductance of the device itself.

\subsection{Change of variables between \texorpdfstring{$I$} \: and \texorpdfstring{$V$}.}
Experimentally, it is usually much easier to bias a device using a current rather than a voltage. To correctly assess our theoretical model, a variable change is needed.
Using two properties of our system
\begin{equation}
    \label{eq:dSdV_to_dSdI_renorm}
    \frac{\partial S}{\partial V_s}\Bigr|_r = \frac{\partial S}{\partial I}\Bigr|_r \frac{\partial I}{\partial V_s}\Bigr|_r
\end{equation}
    and
\begin{align}
    \label{eq:dSdR-V_to_I}
    \frac{\partial S}{\partial r}\Bigr|_{V_s} &=  \frac{\partial S}{\partial r}\Bigr|_I  + \frac{\partial S}{\partial I}\Bigr|_r \frac{\partial I}{\partial r}\Bigr|_{V_s} \nonumber\\
    &= \frac{\partial S}{\partial r}\Bigr|_I  - \frac{\partial S}{\partial I}\Bigr|_r \frac{\partial V_s}{\partial r}\Bigr|_I \frac{\partial I}{\partial V_s}\Bigr|_r,
\end{align}
and after a tedious but straightforward calculation, we arrive at equations 5 and 6 of the main paper
\begin{equation}
    \label{eq:renorm_dVdR}
    \frac{\partial V(I,r)}{\partial r}\Bigr|_I = \frac{1}{2}\frac{\partial S(I,r)}{\partial I}\Bigr|_r,
\end{equation}
\begin{equation}
    \label{eq:renorm_dSdR}
    (R + r) \frac{\partial S(I,r)}{\partial r}\Bigr|_I = -2S + \frac{1}{2} \left(\frac{\partial S(I,r)}{\partial I}\Bigr|_r\right)^2,
\end{equation}
with $R = \frac{\partial V}{\partial I}\Bigr|_r$ the resistance of the device.

\bibliography{SupMat}